\documentclass[12pt]{article}
\usepackage{graphicx}

\newsavebox{\sboxpubnumber}
\newsavebox{\sboxpubdate}
\newcommand{\pubdate}[1]{\begin{lrbox}{\sboxpubdate}{#1}\end{lrbox}}

\newcommand{\Title}[1]{\begin{center} {\Large #1 } \end{center}}
\newcommand{\Author}[1]{\begin{center}{ \sc #1} \end{center}}
\newcommand{\Address}[1]{\begin{center}{ \it #1} \end{center}}

\newenvironment{Abstract}{\begin{quotation}  }{\end{quotation}}
\newenvironment{Presented}{\begin{quotation} \begin{center}
             PRESENTED AT\end{center}\bigskip
      \begin{center}\begin{large}}{\end{large}\end{center}
      \end{quotation}}
\newcommand{\Acknowledgements}{\bigskip  \bigskip \begin{center} \begin{large}
             \bf ACKNOWLEDGEMENTS \end{large}\end{center}}

\begin{document}

\begin{titlepage}
\pubdate{\today}                    

\vfill
\Title{Primordial magnetic fields and CP violation in the sky}
\vfill
\Author{Tanmay Vachaspati\footnote{Work supported in part by
               the Department of Energy, USA.}}
\Address{Physics Department, Case Western Reserve University\\
         Cleveland, OH 44106, USA.}
\vfill
\begin{Abstract}
It has been argued that electroweak baryogenesis generates
a left-handed primordial magnetic field. The helicity
density of the primordial magnetic field today is estimated
to be $\sim 10^2 n_b$ where $n_b\sim 10^{-6}$ /cm$^3$ is
the present cosmological baryon number density. With certain
assumptions about the inverse cascade the field strength at 
recombination is $\sim 10^{-13}$ G on a coherence scale 
$\sim 10^{-4}$ pc. Here I discuss the various assumptions 
made in obtaining these estimates and the concomitant open issues.
\end{Abstract}
\vfill
\begin{Presented}
    COSMO-01 \\
    Rovaniemi, Finland, \\
    August 29 -- September 4, 2001
\end{Presented}
\vfill
\end{titlepage}
\def\thefootnote{\fnsymbol{footnote}}
\setcounter{footnote}{0}


The observed galactic and cluster magnetic fields 
of strength $\sim 10^{-6}$ Gauss
could be due to amplification of a primordial seed
field. Hence it is of interest to find early universe
scenarios that can lead to primordial seed fields.
If the galactic dynamo is maximally efficient, 
the seed field strength should be larger than 
$10^{-21}$ Gauss at the proto-galaxy stage of structure
formation. Another characteristic of the observed magnetic 
fields is their large coherence scale. The ``homogeneous'' 
component of the galactic magnetic field has a typical 
coherence scale $\sim 1$ kpc. Assuming that the
coherence scale cannot grow during galaxy evolution,
the coherence scale of the seed field should also be 
$\sim 1$ kpc at the present epoch.

\section{Magnetic field production}

In Ref. \cite{Vac01} I pointed out that electroweak
baryogenesis would imply a primordial magnetic field
(also see \cite{Cor97}).
The relationship between baryogenesis and magnetic fields
follows once it is realized that baryon number violation
proceeds via twisted non-Abelian gauge fields 
such as the sphaleron or electroweak string configurations. 
These gauge field configurations are the intermediate step 
in processes that change baryon number. The present baryonic
excess is produced as these configurations decay. The
argument in Ref. \cite{Vac01} is that the decay also produces 
an electromagnetic magnetic field, ${\bf B}$, that carries 
some of the twist in the non-Abelian gauge fields making up 
the configuration.
An estimate of the helicity (${\cal H}$) of ${\bf B}$ yields
\begin{equation}
{\cal H} \equiv {1\over V} \int_V d^3 x {\bf A}\cdot {\bf B}
         \sim - {{n_b} \over {\alpha}}
\label{helestimate}
\end{equation}
where $\alpha =1/137$ is the fine structure constant and
$n_b$ is the baryon density of the universe. The sign of the
helicity density is determined by the sign of the baryon
number. Since we observe baryons and not anti-baryons, the
primordial magnetic field has negative helicity density
{\it i.e.} it is left-handed everywhere.

The estimate in eq. (\ref{helestimate}) is based on a specific
decay channel of a certain configuration of electroweak strings,
namely, two linked loops of Z-string. The estimate in 
eq. (\ref{helestimate}) assumes
that other decay channels also lead to roughly the same helicity.
Testing this assumption is the first open issue.

\smallskip

\noindent {\bf Issue 1:} {\it Can one quantify more accurately
the ${\bf B}$ helicity produced by baryon number violating 
processes?}

\smallskip

In principle one could imagine watching a sphaleron decay on a 
computer and then tracking the ${\bf B}$ helicity. This
would provide a check of the above estimate for another decay
channel and would be very valuable. However, the sphaleron would
first have to be constructed for non-zero weak mixing angle
($\theta_w$), perturbed, and then studied. This is a non-trivial
numerical task since one would like to evolve all of the
electroweak bosonic degrees of freedom. Besides, for the purpose
of the present application, we are not interested in the detailed
dynamics of all the bosonic degrees of freedom. We only want to
know how the twist (Chern-Simons number) in the sphaleron gauge 
fields gets transferred to the ${\bf B}$ field helicity. I have
a feeling that it should be possible to extract the latter without
solving for the whole evolution. Whether this can be done is a
very interesting and important question.

\section{Magnetic field evolution}

Once helical magnetic fields are produced, the evolution is
described by the MHD equations in an expanding spacetime.
In the case of a flat geometry, it can be shown that the
field variables can be rescaled so that the evolution is 
given by the MHD equations in a non-expanding spacetime 
\cite{BraEnqOle96}. The initial ${\bf B}$ field will be
produced on very small length scales. How does such a 
magnetic field evolve?

A magnetic field coherent only on very small length scales 
will simply dissipate. This follows from the MHD equation
\begin{equation}
{{\partial {\bf B}} \over {\partial t}} =  
           {\bf \nabla}\times ({\bf v}\times {\bf B})
         + {1\over {4\pi \sigma_c}} {\bf \nabla}^2 {\bf B}
\label{mhdequation}
\end{equation}
where ${\bf v}$ is the plasma velocity and $\sigma_c$ the
electrical conductivity. The second term on the right-hand
side causes dissipation and is proportional to $1/L^2$ where
$L$ is the coherence scale of the field. Therefore when
$L$ is very small the second term greatly dominates 
the first term on the right-hand side and the field decays
exponentially fast. In the opposite situation, where the
first term greatly dominates the second term (i.e. for large
coherence scales and velocity flows) it can be shown that
the magnetic field is frozen-in the plasma and simply gets
dragged with the flow of the plasma. The condition under
which the first term dominates over the second term is that
the magnetic Reynold's number ${\cal R}_M$ be large:
\begin{equation}
{\cal R}_M \equiv 4\pi \sigma_c L v >> 1 \ 
               \Rightarrow \ v >> {{e^4}\over {4\pi}}
\label{largeRM}
\end{equation}
This raises the second open issue.

\smallskip

\noindent {\bf Issue 2:}
{\it What are the plasma
velocities implied by electroweak baryogenesis? Are they
large enough to satisfy the condition in eq. (\ref{largeRM})?}

\smallskip

There are several different scenarios for successful electroweak
baryogenesis. All of them satisfy the Sakharov condition that
thermal equilibrium should be violated during the epoch of
baryogenesis. This is usually achieved by considering a strongly
first order phase transition with typical bubble wall velocities
$\sim 0.1$ (for a review see \cite{Tro99}). In such scenarios the 
fluid velocities are clearly large enough and the first term in
the MHD equation is dominant; yet the second term cannot be
neglected. This brings us to issue 2 -- a rather general
issue involving cosmological phase transitions and MHD.

\smallskip

\noindent {\bf Issue 3:}
{\it How does the magnetic field helicity evolve during and 
after the electroweak phase transition?}

\smallskip

There are some hints regarding this issue that are already
present in the MHD literature in connection with the ``reversed
magnetic pinch''. If a plasma is confined to a cylindrical 
volume and an axial current is applied together with a uniform
axial magnetic field, it is found that the plasma relaxes
in such a way that the final magnetic field at the surface of
the cylinder is anti-parallel to that on the cylinder's axis.
The phenomenon was explained by J.B. Taylor \cite{Tay75} by
hypothesizing that magnetic helicity in an infinite volume
($V\rightarrow \infty$ in eq. (\ref{helestimate}))
is conserved even when the second term in the MHD equation is non-zero
provided it is sub-dominant as compared to the first term. If we adopt
this hypothesis, it would imply that, regardless of the details
of the evolution during the electroweak phase transition, the
global magnetic helicity will be conserved. This is a very
important constraint on the evolution and, as we shall see, will
eventually allow us to estimate the magnetic field strength.

\section{Inverse cascade}

The coherence scale of the magnetic field generated at the 
electroweak phase transition is microphysical. If only the
Hubble expansion were to stretch the coherence scale, this
would amount to a coherence scale of about 1 centimeter today.
However, astrophysical magnetic fields are coherent on kpc 
scales. Is there a way to bridge this enormous gap?

There is evidence in the MHD literature that under certain
circumstances a magnetic field can ``inverse cascade''. A
direct cascade is when energy is transferred from larger
to smaller scales, where it is ultimately dissipated. An
inverse cascade is the transfer of energy from short to
long wavelengths. It is believed that the inverse cascade
can occur provided the plasma is turbulent and the magnetic
field is helical. As discussed above, both these ingredients
are present during the electroweak phase transition. Hence
we can expect an inverse cascade that will stretch the
coherence scale of the magnetic field.

The growth of the coherence scale due to Hubble expansion can 
be factored in
quite trivially and so let us consider MHD in a non-expanding 
spacetime. If $L(t)$ is the time-dependent coherence scale, 
we can write
\begin{equation}
L(t) = L(t_i) \left ( {t \over {t_i}} \right ) ^p
\label{inversecascade}
\end{equation}
where, $t_i$ is some initial time,
and $p$ is the inverse cascade exponent. At present,
there is some debate over the value of $p$. 
Renormalization group calculations indicate that
$p=0$ \cite{BerHoc01}, analytical analyses point to
$p=2/3$ \cite{Frietal75}, while numerical simulations
give $p\sim 1/2$ \cite{BisMul99,Chretal00} The value of $p$
is quite important and the estimates in \cite{Vac01}
take $p=2/3$. This brings us to the next issue.

\smallskip

\noindent {\bf Issue 4:}
{\it What is the value of $p$ and how does it affect the
cosmic evolution of the primordial magnetic field?}

\smallskip

This issue is further complicated by the occurrence of
certain cosmic episodes. For example, when electron-positron 
annihilation takes place, the electrical 
conductivity of the cosmological plasma suddenly drops
and the second term in the MHD equation gains importance.
If $p=1/2$, it turns out that the coherence scale of the 
magnetic field has not grown sufficiently that it can
survive the larger dissipation in the post $e^+e^-$ era. 
In this case, only certain
large wavelength modes of the magnetic field can survive.
The field strength due to these modes will be smaller. However
the analysis of Ref. \cite{Vac01} does not yield the spectrum 
of the generated magnetic field and so we cannot estimate this
field strength. This is issue 5.

\smallskip

\noindent {\bf Issue 5:}
{\it What is the spectrum of the magnetic field 
resulting from electroweak baryogenesis?}

\smallskip

Another cosmic episode that may have an effect on the primordial
helical magnetic field is the QCD phase transition. Depending on 
the nature of the phase transition, there may be turbulence in the
plasma and energy may be injected into the helical magnetic field. 
This would be worth investigating further together with the 
impact of other cosmic events on helical magnetic fields. 

\smallskip

\noindent {\bf Issue 6:}
{\it What effect does the QCD phase transition and other cosmic
events have on pre-existing helical magnetic fields?}

\smallskip

Assuming $p=2/3$, it is easy to estimate the coherence scale
of the magnetic field. This turns out to be
$\sim 10^{14}$ cms at recombination. Once we have the coherence 
scale, we can use
the conservation of magnetic helicity, $\sim B^2 L \sim n_b/\alpha$, 
to estimate the field strength: $B(t_{rec}) \sim 10^{-13}$ Gauss.

The coherence scale corresponds to $\sim 0.1$ pc today and is 
smaller than the observed kpc scale by $10^4$. However, the field
strength is well within what a galactic dynamo can (in principle)
amplify to produce the presently observed field strengths 
of $\sim 10^{-6}$ Gauss.
The primordial field will certainly have some power on kpc scales
but at present we cannot say if it will be sufficient for the dynamo.
That will depend on the spectrum of the field produced during 
baryogenesis (Issue 5).

\section{CP violation in the sky}

Finally I would like to discuss the prospects for observing
a helical magnetic field. Several papers have discussed the
observation of non-helical magnetic fields using the cosmic
microwave background radiation (CMBR) but have missed discussing 
helical fields. The reason is that they
have taken the correlators of the magnetic field
Fourier coefficients to satisfy
\begin{equation}
\langle b_i^* ({\bf k}) b_j ({\bf k}) \rangle
\propto 
\left ( \delta_{ij} - {{k_i k_j} \over {k^2}} \right )
\label{bbcorr}
\end{equation}
where ${\bf k}$ labels Fourier modes, and $\delta_{ij}$ is the 
Kronecker delta function. Note that the right-hand side is
symmetric in $i,j$ and this means that the magnetic field
is non-helical, that is 
$\langle {\bf B}\cdot {\bf \nabla}\times {\bf B} \rangle = 0$. 
Hence, we are
interested in the case when the above correlator has an extra
piece that is anti-symmetric in $i,j$, namely proportional
to $\epsilon_{ijl} k_l$.

The way that a non-helical magnetic field shows up in the CMBR
is that it exerts a Lorentz force on the plasma and causes fluid
flow. The additional velocity of the plasma can be detected in
the photons due to the Doppler effect. This scheme fails for
helical fields since the helical part (the anti-symmetric 
contribution to the correlator in eq. (\ref{bbcorr})) is 
``force-free'' {\it i.e.} does not exert any Lorentz force 
on the plasma! There is a
higher order effect though which is discussed in \cite{PogVacWin01},
due to the different scattering cross-sections of photons off
of electrons and protons. The effect causes non-vanishing correlations
between the CMBR temperature and odd parity polarization
($C_l^{TB}$) and also different types of polarization
($C_l^{EB}$). However the magnitude of these correlators is tiny 
and hence it does not seem likely that the helicity of the
magnetic field can be detected in the CMBR. However, there 
might be other effects that have not been considered and this
is the next issue.

\smallskip

\noindent {\bf Issue 7:}
{\it Can helical magnetic fields be detected in the CMBR?
Or more generally, can the helicity of a cosmological/astrophysical
magnetic field be detected by any means?}

\smallskip

The detection of a helical cosmic magnetic field would be
a direct detection of CP violation in cosmology. In principle, 
the observation of such magnetic fields can lead to a probe of 
the electroweak phase transition and the intervening cosmic history. 
This would be in addition to providing a better understanding of
the observed magnetic fields in galaxies and clusters of galaxies.


\Acknowledgements

I am grateful to George Field, Mark Hindmarsh, Levon Pogosian, 
Kandu Subramanian, and Serge Winitzki for discussions.
This work was supported in part by the Department of Energy, USA.


\begin{thebibliography}{99}


\bibitem{Vac01}
T. Vachaspati, to be published in Phys. Rev. Lett.; 
astro-ph/0101261.

\bibitem{Cor97}
J.M. Cornwall, Phys. Rev. {\bf D56} (1997). 6146. 

\bibitem{BraEnqOle96}
A. Brandenburg, K. Enqvist and P. Olesen,
Phys. Rev. {\bf D54} (1996) 1291-1300.

\bibitem{Tro99}
M. Trodden,
Rev. Mod. Phys. {\bf 71} (1999) 1463-1500.

\bibitem{Tay75}
J.B. Taylor,
Phys. Rev. Lett. {\bf 33} (1974) 1139.

\bibitem{BerHoc01}
A. Berera and D. Hochberg, cond-mat/0103447 (2001).

\bibitem{Frietal75}
U. Frisch, A. Pouquet, J. Leorat and A. Mazure,
J. Fluid Mech. {\bf 68} (1975) 769;
A. Pouquet, U. Frisch and J. Leorat,
J. Fluid Mech. {\bf 77} (1976) 321.

\bibitem{BisMul99}
D. Biskamp and W.C. Muller, 
Phys. Rev. Lett. {\bf 83} (1999) 2195;
W.C. Muller and D. Biskamp, 
Phys. Rev. Lett. {\bf 84} (2000) 475.

\bibitem{Chretal00}
M. Christensson, M. Hindmarsh and A. Brandenburg, 
astro-ph/0011321 (2000).

\bibitem{PogVacWin01}
L. Pogosian, T. Vachaspati and S. Winitzki, 
in preparation (2001).


\end{thebibliography}
\end{document}